\documentstyle[floats,aps,prl,epsf,twocolumn]{revtex}

\begin{document}

\wideabs{

\title{NMR evidence for a "generalized spin-Peierls transition" in the high
magnetic field phase of the spin-ladder
Cu$_{2}$(C$_{5}$N$_{2}$H$_{12}$)$_{2}$Cl$_{4}$ }

\author{H. Mayaffre$^1$, M. Horvati\'c$^2$, C. Berthier$^{1,2}$, M.-H. Julien$^1$,
P. S\'egransan$^1$, L. L\'evy$^{2,3}$, and O. Piovesana$^4$}

\address{
$^{1}$Laboratoire de Spectrom\'etrie Physique, Universit\'e J.
Fourier, BP87, F-38402 St. Martin d'H\`eres, France\\
$^{2}$Grenoble High Magnetic Field Laboratory, CNRS and MPI-FKF,
BP 166, F-38042 Grenoble Cedex 09, France\\ $^{3}$Institut
Universitaire de France, Universit\'e J. Fourier, BP41, F-38402
St. Martin d'H\`eres, France\\ $^{4}$Dipartimento di Chimica,
Universit\'a di Perugia, I-06100 Perugia, Italy}

\date{\today}

\maketitle


\begin{abstract}
The magnetic field-induced 3D ordered phase of the two-leg
spin-ladder Cu$_{2}$(C$_{5}$N$_{2}$H$_{12}$)$_{2}$Cl$_{4}$ has
been probed through measurements of $^1$H NMR spectra and $1/T_1$
in the temperature range 70~mK - 1.2~K. The second order
transition line $T_c(H)$ has been determined between $H_{c1} =$
7.52~T and $H_{c2} =$~13 T and varies as $(H-H_{c1})^{2/3}$ close
to $H_{c1}$. From the observation of anomalous shifts and a
crossover in $1/T_1$ above $T_c$, the mechanism of the 3D
transition is argued to be magnetoelastic, involving a
displacement of the protons along the longitudinal exchange
($J_{\parallel}$) path.

\end{abstract}


}

\narrowtext

Two-leg $S$=1/2 ladders are 1D objects formed by two
antiferromagnetically (AF) coupled Heisenberg spin chains. In zero
external magnetic field, their ground state is a collective
singlet state ($S$ = 0), separated by a gap $\Delta$ from the
first excited states which are triplets ($S=1$) \cite{review}. As
a consequence, the spin-spin correlations remain of short range
even when $T\rightarrow 0$, in spite of the strong interactions.
There is currently considerable interest in these systems, often
named {\it spin-liquids}, since the short-range singlet
correlations of the ground state are believed to lead to
superconducting correlations when mobile charges are added
\cite{review}.

The fascinating properties of spin-liquids can also be revealed
through the effect of a magnetic field $H$. This can be described
in four steps: (1)~For $H\neq 0$, the gap is reduced as $\Delta(H)
= \Delta -g\mu _{B}H$. (2)~At the so-called quantum critical point
$H=H_{c1}=\Delta/g\mu _{B}H$, the spin-gap vanishes. At $T$ = 0,
this defines a (quantum) phase transition between gapped singlet
and gapless magnetic phases. (3)~For $H>H_{c1}$, the gapless spin
system still exhibits 1D behavior at finite temperature, but as
$T$ is reduced the magnetic correlation length and the spin-spin
correlation functions now diverge (Luttinger liquid behaviour).
This behaviour can be observed up to a saturation field $H_{c2}$
where all spins are polarized by $H$. (4)~For $H_{c1}<H<H_{c2}$,
the transverse coupling $J_t$ between ladders should drive the
system towards a 3D magnetic ordering at low $T$. The nature of
the 3D phase, in the vicinity of the two quantum critical points,
is expected to be highly unconventional.

Points (1-3) were previously observed in NMR studies
\cite{ChaboussantPRL1,ChaboussantPRL2,ChaboussantEPJ} of the
spin-ladder Cu$_{2}$(C$_{5}$N$_{2}$H$_{12}$)$_{2}$Cl$_{4}$
\cite{Chiari} in which the low values of the AF exchange coupling
 (between spins 1/2 on Cu$^{2+}$ ions) along the legs
($J_{\parallel}\simeq $ 3 K) and along the rungs ($J_{\bot }\simeq
$ 13 K), lead to experimentally accessible values of $H_{c1}$
($\simeq $ 7.5 T) and $H_{c2}$ ($\simeq $ 13.5 T)
\cite{ChaboussantPRB}. As to point (4), specific heat measurements
\cite{Hammar,Calemczuk,Hagiwara} in the field range 7-12 T have
indeed revealed a phase transition towards a 3D ordered phase for
$T<1$~K. However, no microscopic experimental insight has been
reported so far, although this phase currently generates a large
interest
\cite{ChaboussantEPJ,Hammar,Calemczuk,Hagiwara,giamtsvelik,Nagaosa,Mila,Wessel}.

In this Letter, we present a $^1$H NMR study of
Cu$_{2}$(C$_{5}$N$_{2}$H$_{12}$)$_{2}$Cl$_{4}$ in the field range
7.5-14 T, including the $T$-dependence (in the range 70 mK-1.2 K)
of the lineshape and of the nuclear spin-lattice relaxation rate
$1/T_{1}$. From the splitting of NMR lines, we define the
transition line $T_c(H)$ below which 3D ordering occurs. In
addition, we observe through $1/T_{1}$ a drastic change in the
low-energy spin excitations below $\simeq$1.3 K, which is above
$T_c$. This behavior is correlated with anomalous shift of some
$^1$H lines, which we attribute to the displacement of protons
involved in the exchange path along the legs of the ladder. This
is argued to demonstrate the magneto-elastic nature of the
transition, which is in some way analogous to the incommensurate
magnetic phase of spin-Peierls systems.

\begin{figure}[tbp]
\centerline{\epsfxsize=80mm \epsfbox{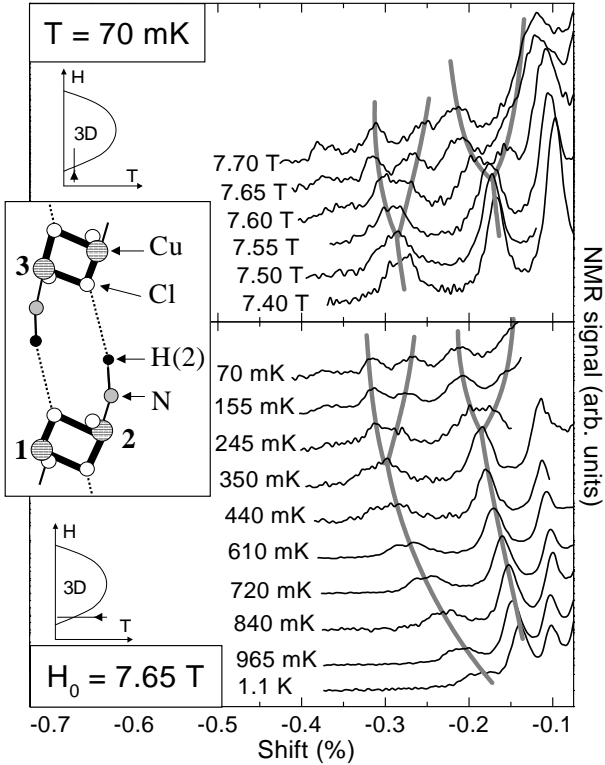}} \caption{Upper
panel: low frequency side of $^1$H spectra recorded at fixed
values of $H$ at $T$ = 70 mK. The onset of the transition at $H=
7.55$~T is revealed by the splitting of the lines. Lower panel:
low frequency side of $^1$H spectrum at various $T$ for $H$= 7.65
T; the transition occurs between 350 and 245 mK.} \label{Fig1}
\end{figure}

\begin{figure}[tbp]
\centerline{\epsfxsize=80mm \epsfbox{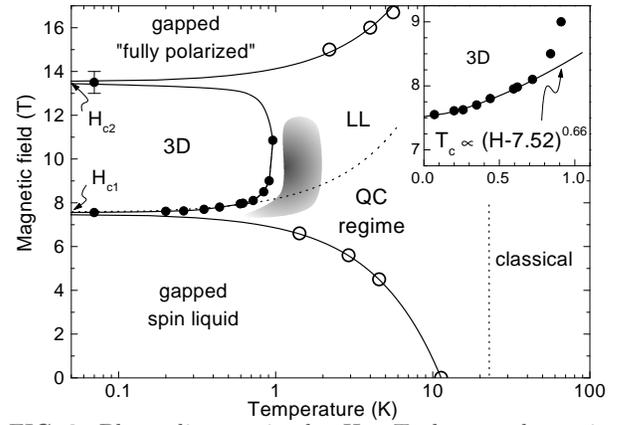}}
\caption{Phase diagram in the $H-T$ plane as determined from NMR
experiments. The shaded area corresponds to the $T$-range in which
proton motion is observed (see text). The inset shows the
variation of $T_c(H)$ in the vicinity of $H_{c1}$, the solid line
is a fit to $\propto(H-H_{c1})^{2/3}$.} \label{Fig2}
\end{figure}

Experiments have been performed on a single crystal placed inside
the mixing chamber of a $^{3}$He-$^{4}$He dilution refrigerator,
the $b$-axis of the crystal being parallel to $H$. In this
orientation, the number of inequivalent $^1$H sites in the crystal
is reduced to 24. All sites experience different hyperfine fields
through their dipolar coupling to the electronic spins localized
at the Cu sites \cite{remark0}. For large polarization of the
electronic moments, these couplings lead to $^{1}$H spectra
extending over several MHz, which have been recorded at fixed
field by sweeping the frequency. In Fig. \ref{Fig1} is shown the
low frequency part of such spectra recorded at 70 mK at various
values of $H$. One clearly observes a splitting of all lines
starting at $H=7.55$~T and increasing for higher $H$ values. This
is the signature of an ordered magnetic phase. Following the
evolution of the spectrum with $H$ for different $T$ values allows
an accurate determination of the transition line $T_{c}(H)$. As
shown in Fig. \ref{Fig2}, $T_{c}$ rapidly increases as a function
of $H-H_{c1}$ and then saturates around 900 mK for $H\geq$ 9 T. In
this range, the transition was determined at fixed value of $H$
and decreasing $T$. Again, we observe a line splitting
(Fig.~\ref{Fig1}), which quickly increases below $T_{c}$ and then
saturates at lower $T$, as expected for an order parameter. The
resulting experimental phase diagram is shown in Fig.~\ref{Fig2}.

There are a few theoretical calculations of the line $T_c(H)$:
Close to $H_{c1}$. Giamarchi and Tsvelik predict a variation as
$(H-H_{c1})^{2/3}$, resulting from the condensation of dilute hard
core bosons \cite{giamtsvelik}. Wessel and Haas rather propose an
$(H-H_{c1})^{1/2}$ variation \cite{Wessel}. As shown in the inset
to Fig. \ref{Fig2}, $T_{c}$ can be well-fitted to $\propto
(H-H_{c1})^{2/3}$, with $H_{c1}=7.52$ T. Note that the prediction
of a first order transition with a jump of magnetization for
values of $H$ close to $H_{c1}$ \cite{Nagaosa} is not observed in
our data even very close to $H_{c1}$. In all cases, each line
splits at $T_c$ keeping the same center of gravity, thus
indicating that the magnetization is continuous with $T$ or $H$,
in agreement with earlier thermodynamic
measurements~\cite{ChaboussantPRB}.

\begin{figure}[tbp]
\centerline{\epsfxsize=80mm \epsfbox{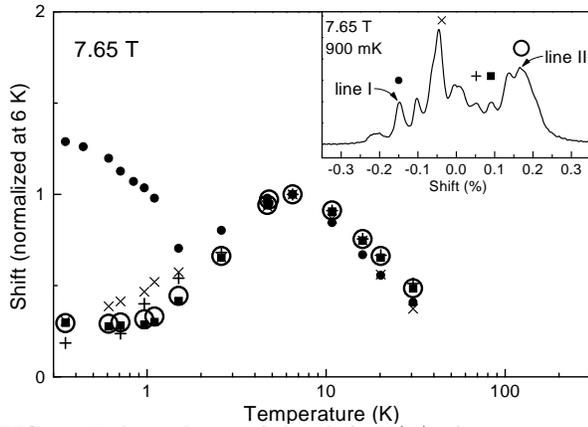}}
\caption{$T$-dependence of the shift $K(T)$ of various proton
lines (the full spectrum is shown in the inset), normalized at
6~K. $K(T)$ scales to the uniform spin susceptibility for all
lines, except for those corresponding to protons H(2) and H(4)
which strongly deviate below $T \simeq$ 1.5~K. This can only be
explained if the position of these protons, hence their hyperfine
field, varies in this $T$-range. } \label{Fig3}
\end{figure}

A careful examination of the whole $^{1}$H spectrum
(Fig.~\ref{Fig3}) {\it above $T_c$} reveals that the
$T$-dependence of the shifts of the two lines at the left side of
the spectrum does not scale with that of other $^1$H lines. These
two lines are assigned to the protons H(2) and H(4)
\cite{ChaboussantPRL1}, which are along the exchange path
$J_{\parallel}$ corresponding to the atom sequence Cu-N-H$\cdots
$Cl-Cu (see inset to Fig.~\ref{Fig1}).
Since the shift of a proton $^{1}$H($i$) is given by $\delta%
h(i)=A(i)\chi_{Cu}$, in which $A(i)$ is its hyperfine field and
$\chi_{Cu}$ the spin susceptibility per Cu atom, the absence of
scaling can only be explained if $A(2)$ and $A(4)$ become
$T$-dependent. This, in turn, can only occur if the distances
H(2)-Cu and H(4)-Cu change \cite{distance}. This is thus clear
evidence that some kind of lattice instability occurs prior to the
magnetic ordering.

Since these protons are along the exchange path corresponding to
$J_{\parallel}$, any modification of the hydrogen bond should
clearly change the magnetic excitation spectrum of the system,
which can be probed by the nuclear spin-lattice relaxation rate
($1/T_1$).

\begin{figure}[tbp]
\centerline{\epsfxsize=80mm \epsfbox{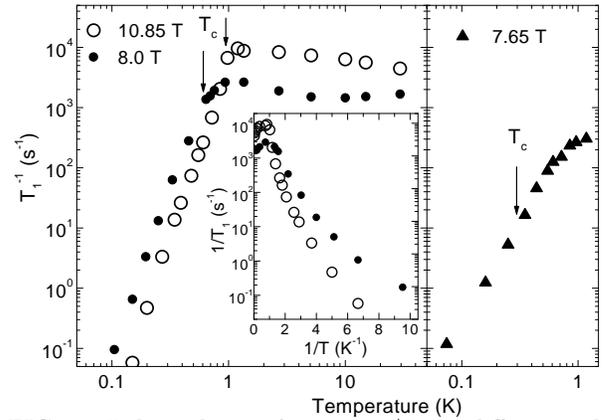}}
\caption{$T$-dependence of proton $1/T_1$ at different values
of~$H$. Left panel: data at $H$ = 8.0~T and 10.85~T, measured on
line II (see inset to Fig.3~). Data above 1.1 K are taken from
\protect\cite{ChaboussantPRL2}. Right panel: data at $H$ = 7.65 T
measured on line~I.} \label{Fig4}
\end{figure}

Such change is indeed observed in $T_1$ data, measured between
1.2~K and 70~mK and reported in Fig.~\ref{Fig4}. There are three
striking features in these data: {\it i}) the huge decrease of
$1/T_{1}$ up to 5 orders of magnitude at $H$= 8.0 T and 10.85 T.
This decrease can be fitted by a power law ($1/T_{1} \propto
T^{-5}$) \cite{remark1}. {\it ii}) the increase of $1/T_1$,
attributed to the divergence of the spin correlation functions
\cite{ChaboussantPRL2}, stops around 1.3 K for all field values
between $H_{c1}$ and $H_{c2}$. {\it iii}) $1/T_1$ starts
decreasing {\it before} the onset of the 3D transition (as
detected by the modification of the lineshape).

We now discuss the possible nature of this 3D ordered ground
state. From a theoretical point of view, the spin-ladder
Hamiltonian can be transformed into an interacting spinless
fermion model through the canonical Jordan-Wigner transformation
\cite{Chitragiam}. In this representation, $H$ acts as the
chemical potential $\mu$, and for $H=H_{c1}$, $\mu$ lies exactly
at the bottom of the band. Increasing $H$ further fills the band
in. Since the value of the Fermi-wave vector $k_{F}$ is set by the
field, $k_F$ is {\it incommensurate} (IC) with the underlying
lattice, except at half filling. Due to the divergence of the spin
susceptibility at 2$k_{F}$, the on-site magnetization of the
ordered phase is also expected to be incommensurate. Between
$H_{c1}$ and $H_{c2}$, and at sufficiently low $T$, the low energy
properties of the system are those of a Luttinger liquid
\cite{Chitragiam}.

In the same field range, the spin-ladder Hamiltonian can also be
approximately mapped onto that of an $XXZ$ $S$=1/2 chain
\cite{ChaboussantEPJ,Mila}. In this latter representation, an
effective spin 1/2 is introduced, whose eigenstates correspond to
the singlet and the lowest state of the triplet on a rung, and the
effective field $H_{\rm eff}$ is equal to $H-(H_{c1}+H_{c2})/2$.
In the following discussion, we shall use either the spinless
fermion or the XXZ language.

It is well known that there are two possibilities to achieve 3D
ordering at finite $T$ for quantum spin chains: a transverse
magnetic coupling $J_t$, leading to some kind of AF order when
$J_t \xi_{\parallel}^2\simeq k_B T$ or a spin-Peierls (SP)
transition in presence of magneto-elastic coupling \cite{Bray}. In
the latter case, a modulation of the lattice occurs, which is
stabilized by the energy gain due to the opening of a gap in the
magnetic excitation spectrum. The 3D character of the transition
arise, in this case, from to the 3D nature of the elastic modes.

The case of a transverse magnetic coupling for an assembly of
ladders has been treated by Giamarchi and Tsvelik
\cite{giamtsvelik}. In their model, the 3D ordering corresponds to
a freezing of the XY degrees of freedom of the triplet states, and
below $T_c$, the local magnetization $M_{z}(R)$ is incommensurate
along the direction of the ladders. A magneto-elastic scenario has
been treated by Nagaosa and Murakami \cite{Nagaosa}, who
considered a modulation of the exchange along the rungs $J_{\bot
}$ and by Calemczuk {\it et al.} \cite{Calemczuk} who found that a
modulation of $J_{\parallel}$ better explains specific heat data
\cite{remark11}. As in the purely magnetic scenario, the local
magnetization $M_{z}(R)$ is IC along the ladder direction, and the
3D ordered phase is in some sense similar to the IC magnetic phase
observed in spin-Peierls systems above their threshold field $H_c$
\cite{Bray}. There is, however, a noticeable difference: in
regular SP compounds, there is a commensurate (dimerized) phase,
which is a collective singlet for $0 < H < H_c$. In the
spin-ladder system, the commensurability occurs for $H
=(H_{c1}+H_{c2})/2$ ({\it i.e.} $H_{\rm eff}\simeq 0$). Any
extension of the commensurability around this $H$ value would
correspond to a plateau in the magnetization (which is not
observed in Cu$_{2}$(C$_{5}$N$_{2}$H$_{12}$)$_{2}$Cl$_{4}$). Along
the same line, the parts of the phase diagram close to $H_{c1}$ or
$H_{c2}$ in the ladder case correspond to a field range close to
the saturation of the magnetization in the case of a regular SP
system.

We now compare our data to the predictions of these different
models. All of them predict an IC modulation of $M_{z}(R)$, giving
rise to an infinite number of inequivalent sites. This should
transform each NMR line into a double horned lineshape
\cite{Blinc}. Because of the high density of $^{1}H$ lines in our
spectra, we cannot distinguish whether each line transforms this
way or simply splits.

At variance with the lineshape, the $T$-dependence of $1/T_{1}$ is
expected to depend strongly on the nature of the ground state. A
purely magnetic ground state \cite{Hammar,giamtsvelik} implies a
divergence of $1/T_{1}$ at the transition, which is not observed
experimentally.

The increase of $1/T_1$ upon cooling, seen on Fig.~4 above $\simeq
1.3$~K, is indeed related to the Luttinger liquid behavior of the
gapless 1D system, as explained in \cite{ChaboussantPRL2}. This
increase cannot be attributed to critical fluctuations linked to
the transition, since it starts too far above $T_c$
\cite{ChaboussantPRL2} and furthermore we now find that it even
stops above $T_c$. This is particularly obvious for the data at
7.65~T (right panel of Fig.~4).

In contrast, in a spin-Peierls like transition, the low-energy
spectral weight of AF fluctuations starts being suppressed even
above the 3D ordering by the coupling to the elastic degrees of
freedom. Hence, there is no divergence, but a rapid decrease of
$1/T_{1}$ due to the opening of a gap. This was experimentally
observed at the transition between the IC high field phase and the
uniform phase in the spin-Peierls compound CuGeO$_{3}$
\cite{Fagot98}. Due to the IC nature of the ground state, the
relaxation rate below $T_c$ should be dominated by the phasons,
which are the standard Goldstone modes of IC phases, so that the
decrease of $1/T_{1}$ is not necessarily thermally activated. This
could be the origin of the apparent power law observed here. As
shown in Fig.~\ref{Fig4}, a faster decrease is observed for
$H=10.85$~T, which corresponds approximately to $H_{\rm
eff}=(H_{c1}+H_{c2})/2$ where commensurability occurs. For
$H=7.65$~T, a field close to $H_{c1}$, the decrease is noticeably
slower. However, for this field value, $1/T_{1}$ was measured on
the line I [protons H(2) and H(4)], while it was measured on the
line II for $H=8.0$~T and 10.85~T (see inset to Fig.~3). As
discussed in Ref.~\cite{ChaboussantPRL1}, the corresponding sites
have different form factors, and do not probe the same linear
combination of the transverse and longitudinal spin-spin
correlation functions. So, we cannot really attribute the weaker
$T$-dependence of $1/T_1$ at 7.65~T to the proximity to $H_{c1}$.

In summary, for approximately the same $T$-range where $1/T_{1}$
decreases, we observe displacements of protons located in the
hydrogen bonding along the exchange path $J_{\parallel}$. These
three features, namely i) absence of $1/T_{1}$ divergence at
$T_c$, ii) decrease of $1/T_{1}$ above $T_c$ and iii) evidence for
proton displacements in the same $T$ range, rule out any model
involving solely a magnetic coupling between the ladders
\cite{Hammar,giamtsvelik}, and strongly support a "generalized
spin-Peierls" scenario. We also found that the field dependence of
$T_{c}$ is consistent with the predictions of a Bose condensation
type of transition ($T_{c}\propto (H-H_{c1})^{2/3}$).

It must be stressed that NMR spectra only tell us about the
time-averaged displacements of these protons. Would the
displacements of protons H(2) and H(4) be purely static, only the
value of $J_{\parallel}$ would change (and thus that of $H_{c1}
\propto J_{\bot}-J_{\parallel}$). The dynamics, evidenced by the
divergence of $1/T_{1}$ in the 1D regime, would not be affected.
To alter the magnetic excitation spectrum, a dynamical modulation
of the position of these protons must be present. In other words,
they have to participate to some phonon mode coupled to the
magnetic excitations and leading to a dynamic modulation of
$J_{\parallel}$. This magneto-elastic coupling appears prior to
the "spin-Peierls transition" at $T_{c}(H)$ and it readily
explains the change in the $T$-dependence of $1/T_{1}$ above
$T_c$. The freezing of this collective mode would finally lead to
a static IC modulation of the position of the protons along the
legs of the ladder.

We note that preliminary data show that anomalous shifts, related
to proton movements, are still observed very close to $H_{c1}$,
where $T_c\rightarrow 0$.  This can be explained by the fact that,
already at $H=H_{c1}$, the system has interest in suppressing the
quantum magnetic fluctuations \cite{hc1} through the
magneto-elastic coupling. However, an ordering cannot be achieved
at finite temperature since there is no finite magnetization along
the individual ladders.

We thank Th. Giamarchi for discussions and P. Van der Linden for
technical help.


\begin{references}
\bibitem{remark0} For those protons which are the closest to the Cu
atoms, some supertransferred hyperfine interaction is very likely
too.
\bibitem{distance} The dipolar coupling with electronic spins varies as
$r^{-3}$. A supertransferred hyperfine field, through the path
Cu-N-H, should vary exponentially with the N-H distance.
\bibitem{remark1}  Although $1/T_{1}$ decreases by five order of magnitude,
this occurs over only one decade in $T$ so that an activated
behaviour below 250 mK cannot be fully excluded (inset to Fig.4~).
\bibitem{remark11} Considering the structure of the ladder, it is
much easier to modify the exchange along the hydrogen bond, than
to modulate the Cu-Cl-Cu bridges.
\bibitem{hc1} At $H=H_{c1}$, the divergence of the spin-spin
correlation functions is seen through $1/T_1\propto T^{-0.5}$
\cite{ChaboussantPRL2}.
\end{references}
\end{document}